\documentclass[final,5p,times,twocolumn,number,sort&compress]{elsarticle}
\usepackage{graphicx,epstopdf,amsmath}
\usepackage{booktabs}
\usepackage{multirow}
\begin{document}
\begin{frontmatter}

\title{Precise Mass Measurements of $A=133$ Isobars with the Canadian Penning Trap: Resolving the $Q_{\beta^-}$ anomaly at $^{133}$Te}

\author[1,2]{A.A. Valverde \corref{cor1}}
\ead{avalverde@anl.gov}

\author[1]{F.G. Kondev}
\author[1,3]{B. Liu}
\author[1,2]{D. Ray \fnref{fn1}}
\author[3]{M. Brodeur}
\author[1]{D.P. Burdette}
\author[1]{N. Callahan}
\author[1,3]{A. Cannon}
\author[1,2]{J.A. Clark}
\author[4]{D.E.M Hoff}
\author[5]{R. Orford}
\author[3]{W.S. Porter}
\author[1,6]{G. Savard}
\author[2]{K.S. Sharma}
\author[1,6]{L. Varriano \fnref{fn2}}

\affiliation[1]{organization={Physics Division, Argonne National Laboratory},
city={Lemont},
state={Illinois},
postcode={60439},
country={USA}
}

\affiliation[2]{organization={Department of Physics and Astronomy, University of Manitoba},
city={Winnipeg},
state={Manitoba},
postcode={R3T 2N2},
country={Canada}
}

\affiliation[3]{organization={Department of Physics and Astronomy, University of Notre Dame},
city={Notre Dame},
state={Indiana},
postcode={46556},
country={USA}
}

\affiliation[4]{organization={Nuclear and Chemical Sciences Division, Lawrence Livermore National Laboratory},
city={Livermore},
state={California},
postcode={94550},
country={USA}
}

\affiliation[5]{organization={Nuclear Science Division, Lawrence Berkeley National Laboratory},
city={Berkeley},
state={California},
postcode={94720},
country={USA}
}

\affiliation[6]{organization={Department of Physics, University of Chicago},
city={Chicago},
state={Illinois},
postcode={60637},
country={USA}
}

\cortext[cor1]{Corresponding author}

\fntext[fn1]{Present address: TRIUMF, Vancouver, BC V6T 2A3, Canada}
\fntext[fn2]{Present address: Center for Experimental Nuclear Physics and Astrophysics, University of Washington, Seattle, WA 98195, USA}

\begin{abstract}
We report precision mass measurements of $^{133}$Sb, $^{133g,m}$Te, and $^{133g,m}$I, produced at CARIBU at Argonne National Laboratory's ATLAS facility and measured using the Canadian Penning Trap mass spectrometer. These masses clarify an anomaly in the $^{133}$Te $\beta$-decay. The masses reported in the 2020 Atomic Mass Evaluation (M. Wang \emph{et al.}, 2021) produce $Q_{\beta^-}(^{133}$Te)=2920(6)~keV; however, the highest-lying $^{133}$I level populated in this decay is observed at $E_i=2935.83(15)$ keV, resulting in an anomalous $Q_{\beta^{-}}^{i}=-16(6)$~keV. Our new measurements give $Q_{\beta^-}(^{133}\text{Te})=2934.8(11)$ keV, a factor of five more precise, yielding $Q{_\beta^i}=-1.0(12)$~keV, a 3$\sigma$ shift from the previous results. This resolves this anomaly, but indicates further anomalies in our understanding of the structure of this isotope.
\end{abstract}

\end{frontmatter}
\section{Introduction}
Mass is a basic property of the atomic nucleus and can be measured with high-precision by various mass-spectrometry techniques~\cite{Yam21}. Nuclear masses give direct access to the binding energy in a given nucleus and to the energy available for a given nuclear reaction. Thus, nuclear masses are important for understanding many nuclear structure and reaction phenomena, including the origin of the elements in the universe.

In radioactive decay, the masses of the parent and daughter nuclei determine the maximum decay energy, the so-called $Q$ value, which restricts the energy window available for the decay to proceed.

\begin{figure}[hb!]
\centering
 \includegraphics[width=\columnwidth]{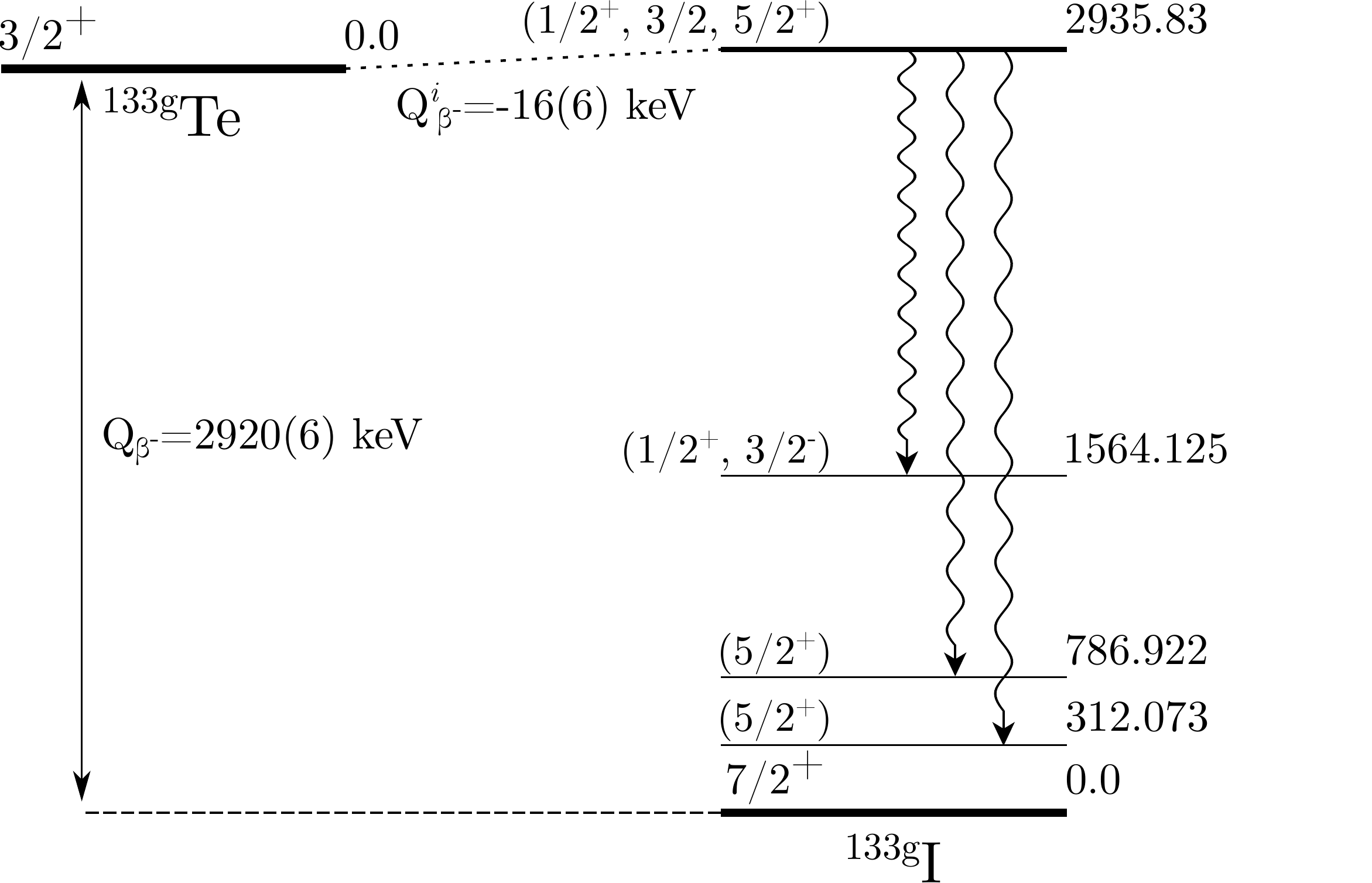}
 \caption{Schematic diagram of $^{133g}$Te $\beta^{-}$-decay through the $E_{i}$=2935.83-keV level in $^{133}$I ~\cite{Khazov11,Hicks83}. The quoted $Q_{\beta^-}$ value is from AME2020~\cite{AME20}, while the spins, parities and excitation energies to the shown levels are from Ref.~\cite{Khazov11}.\label{fig:1Decay}}
\end{figure}

Often, excited energy levels ($E_{i}$) are populated in the daughter nuclei and these $E_{i}$ can be near $Q$. Conservation of energy dictates that for populated levels, $E_{i} \leq Q$. This can result in relatively low effective $Q$ values of $Q^{i} = Q - E_{i}$. An unusual case of such decays occurs in the $\beta^{-}$ decay of the $^{133g}$Te ground state (I$^{\pi}$=3/2$^{+}$, T$_{1/2}$=12.5(3) min~\cite{NUBASE20}). 
According to the 2020 edition of the Atomic Mass Evaluation (AME2020), $Q_{\beta^-}(^{133g}$Te)=2920(6) keV~\cite{AME20}. 
However, the highest-lying level in the daughter nucleus $^{133}$I populated in this decay is observed at $E_{i}$=2935.83(15) keV~\cite{Khazov11,Hicks83}, which results in anomalous $Q_{\beta^{-}}^{i}=-16 (6)$ keV. A schematic diagram of this decay can be seen in Fig. \ref{fig:1Decay}.
Such an nonphysical result must come from either an incorrect decay scheme or from inaccurate mass data used in the determination of $Q_{\beta^-}$. 

The 2935.83(15) keV level is assigned I$^{\pi}=(1/2^{+}, 3/2, 5/2^{+})$ in Ref.~\cite{Khazov11} and it is associated with the decay of the low-spin $^{133g}$Te ground state~\cite{Khazov11,Hicks83}. 
The known $\beta^{-}$-decaying isomer in $^{133m}$Te (I$^{\pi}$=11/2$^{-}$, T$_{1/2}$=55.4 (5) min, E$_{m}$=334.26 (4) keV~\cite{NUBASE20}) decays to relatively high-spin (I=9/2, 11/2 or 13/2)  levels in $^{133}$I where the last one is located at 3051.30 (8) keV~\cite{Walters84,Khazov11}.

The AME2020 $Q_{\beta^-}$ value for $^{133g}$Te is primarily derived from the masses of $^{133g}$Te and $^{133g}$I measured using Penning traps and the time-of-flight ion-cyclotron-resonance (TOF-ICR) technique by Hakala~{\it et al.}~\cite{Hakala12} (93\% significance) and Van~Schelt~{\it et.~al.}~\cite{VanSchelt13} (84\% significance), respectively. 
It should be noted, however, that the mass of $^{133g}$Te reported by Hakala {\it et al.}~\cite{Hakala12} is 38(7) keV smaller than that reported by Van Schelt {\it et al.}~\cite{VanSchelt13}. 
Given the inconsistency between the decay spectroscopy and mass spectrometry data, the disagreement between the different mass measurements (see Table~\ref{tab:mass}), and the improvement in precision in Penning trap mass spectrometry through the rise of the phase imaging ion-cyclotron-resonance (PI-ICR) technique~\cite{Eliseev13} facilitating the separation of isomeric states since these mass measurements were taken, new measurements of both masses are needed to resolve the anomaly.

\section{Experimental Setup and Analysis}
In this Letter, we report new mass measurements for three neutron-rich $A=133$ isobars produced at the CAlifornium Rare Isotope Breeder Upgrade (CARIBU)~\cite{Savard08,Savard11} at Argonne National Laboratory's ATLAS facility and then measured using the Canadian Penning Trap mass spectrometer (CPT)~\cite{Savard01}. 
Beam production at CARIBU begins with the spontaneous fissioning of a $\sim0.5$ Ci $^{252}\text{Cf}$ source; fission fragments were slowed with a thin gold foil and then stopped in a helium-filled large-volume gas catcher. 
Preliminary separation is accomplished using a high-resolution coupled-magnetic-dipole mass separator~\cite{Davids08}, which for this experiment were tuned for $A/q=133/1^{+}$, producing a beam with $^{133}\text{Sb}^+$,  $^{133g,m}\text{Te}^+$, and  $^{133g,m}I^+$ as the primary components. 
This continuous beam from the gas catcher was injected into a radiofrequency quadrupole (RFQ) cooler-buncher where the emittance and energy spread of the ions was reduced through collisions with moderate-pressure high-purity helium gas. The ions were then accumulated and released in bunches every 50 ms. 
These bunches were injected into a multi-reflection time-of-flight mass separator (MR-TOF)~\cite{Hirsh16} for approximately 14.5 ms, allowing for the separation of individual species of the ions of interest from isobaric contaminants with a resolving power $\frac{m}{\Delta m} > 10^5$ using a Bradbury-Nielsen gate after the MR-TOF exit. 
Stable beams of $^{133}\text{Cs}^+$, which was used as the calibrant for all three isotopes of interest, were produced using the CPT's Stable Ion Source, a thermal alkali source located upstream of the CPT tower. 
Both of these low-energy beams were delivered to the CPT experimental setup, where they are first captured and further cooled in a linear RFQ trap before being injected into the CPT. 

\begin{figure}[t!]
\centering
 \includegraphics[width=\columnwidth]{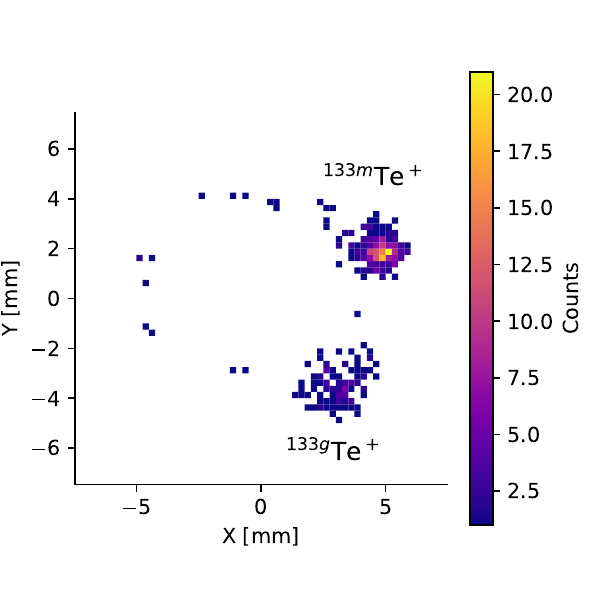}
 \caption{A histogram of detected ion locations for a sample $^{133g}\text{Te}^+$ final phase measurement with phase accumulation time $t_{\text{acc}}=450.493$ ms. The locations of the two species captured in the trap, which separate due to differing mass-based phase accumulations, are labeled. \label{fig:2Spots}}
\end{figure}

Once captured in the Penning trap, the PI-ICR technique~\cite{Eliseev13} was used to conduct a direct mass measurement. 
A detailed description of the identification and measurement scheme as implemented at the CPT can be found in Ref.~\cite{Orford20}, but in brief, the cyclotron frequency ($\nu_c$) of an ion is determined through a simultaneous measurement of the reduced cyclotron frequency and the magnetron frequency.
This is done through two classes of measurement for a fixed total time in the trap. First, a measurement of a reference phase where only magnetron motion is accumulated during the time in the trap. Subsequently, a final phase measurement is conducted, where reduced cyclotron motion is allowed to accumulate for some time $t_{\text{acc}}$ and then the motion is converted to magnetron motion which then accumulates for the rest of the time in the trap. The cyclotron frequency can then be determined by:
\begin{equation}
\nu_c=\frac{\phi_{\text{tot}}}{2\pi t_{\text{acc}}}=\frac{\phi_c+2\pi N}{2\pi t_{\text{acc}}},
\end{equation}
where $\phi_c$ is the measured difference between the final and reference phase, and $N$ is the number of complete revolutions the ion undergoes in $t_{\text{acc}}$. 
After several shorter $t_{\text{acc}}$, from single to hundreds of ms, are used to identify all beam components, including isomeric states, a longer $t_{\text{acc}}$ measurement is done to produce a final measurement, following the techniques described by Ref. ~\cite{Orford20} to minimize potential systematic uncertainties. 
A sample result for a final phase measurement for $t_{\text{acc}}=450.493$ ms of $^{133g}\text{Te}^+$ with all the visible beam components identified can be seen in Fig.~\ref{fig:2Spots}; spots were then clustered and analyzed using a Gaussian mixture model code~\cite{Weber22}.

As the ions of interest and the calibrant were isobaric, this eliminates most potential sources of systematic uncertainties. 
One remaining source of error is in the reference phase measurement, from the small but different accumulation of mass-dependent phase during the excitation period for the ground and isomeric states. This is corrected by following the method described in Ref.~\cite{Orford20}, following an iterative process where this correction is recalculated based on the newly-calculated cyclotron frequency until the change in the correction is more than an order of magnitude smaller than the statistical uncertainty; for these data, this correction was smaller that a part in $10^{10}$. 

%%%%%%%%%%%%%%%%%%%%%%%%%%%%%%%%%%%%%%%%%%%%%%%%%%%%%%%%%%%%%%%%%%%%%%%%%%%%%%%%%%%%%%%%
\begin{table*}[t!]
 \caption{\label{tab:mass} Spins and parities (I$^{\pi}$), half-lives (T$_{1/2}$), cyclotron frequency ratios ($R$), and mass excesses (\emph{ME}) for the ground and isomeric states of nuclides ($X$) measured in the present work, compared to literature values.}
%\begin{ruledtabular}
\begin{tabular}{c c c c c c c c}
\hline
       & & & \multicolumn{2}{c}{Present work}  & \multicolumn{3}{c}{Literature values, [keV]} \\
\cmidrule(lr){4-5}\cmidrule(lr){6-8}
Nuclide            & I$^{\pi}$$^{a)}$ & T$_{1/2}$$^{a)}$ & $R=\bar{\nu}_c(^{133}\text{Cs}^+)/\bar{\nu}_c(^{133}\text{X}^+)$            & \emph{ME} [keV]         & \emph{ME}$^{a,b)}$ & \emph{ME}$^{c)}$ & \emph{ME}$^{e)}$     \\ \hline
$^{133g}\text{Sb}$ &(7/2$^{+}$)        & 2.34 (5) {\it m}  & 1.000073910 (9) & -78 920.9 (11)& -78 924 (3)    & -78 921.3 (76)          & -78 921.0 (40) \\
$^{133g}\text{Te}$ &3/2$^{+}$\#        & 12.5 (3) {\it m}  & 1.000041494 (6) & -82 934.0 (7) & -82 937.1 (21) & -82 899.8 (65)          & -82 938.2 (22) \\
$^{133m}\text{Te}$ &(11/2$^{-}$)       & 55.4 (5) {\it m}  & 1.000044209 (6) & -82 597.8 (7) & -82 602.8 (21) & -82 599.1 (24)$^{d)}$   & -82 595.8 (24)$^{f)}$ \\
$^{133g}\text{I}$  &7/2$^{+}$          & 20.83 (8) {\it h} & 1.000017788 (7) & -85 868.7 (9) & -85 857 (6)    & -85 858.2 (64)          &                 \\
$^{133m}\text{I}$  &(19/2$^{-}$)       & 9 (2) {\it s}     & 1.000031003 (46)& -84 232.8 (57)& -84 223 (6)    &                &                 \\
\hline
\end{tabular}
%\end{ruledtabular}
$^{a)}$NUBASE2020~\cite{NUBASE20}.; 
$^{b)}$AME2020~\cite{AME20}.;
$^{c)}$Van Schelt \emph{et al.}~\cite{VanSchelt13}.;
$^{d)}$Van Schelt~\cite{VanSchelt12}.;
$^{e)}$Hakala \emph{et al.}~\cite{Hakala12}.;
$^{f)}$Kankainen \emph{et al.}~\cite{Kankainen13}.\\
\end{table*}
%%%%%%%%%%%%%%%%%%%%%%%%%%%%%%%%%%%%%%%%%%%%%%%%%%%%%%%%%%%%%%%%%%%%%%%%%%%%%%%%%%%%%%%%

\begin{figure}[t!]
\centering
 \includegraphics[width=\columnwidth]{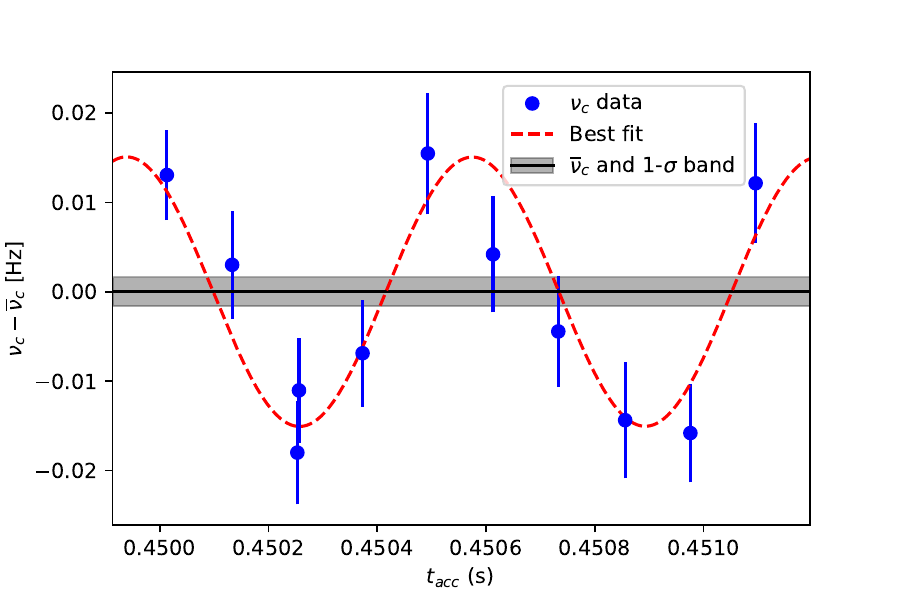}
 \caption{Measured $\nu_c$ values for $^{133g}\text{Te}^+$ at eleven distinct $t_{\text{acc}}$ values between 450.013 ms and 451.096 ms. The dashed line represents a fit of the model described in Ref. \cite{Orford20} to the data, and the solid line and bar the true $\bar{\nu}_c$.  \label{fig:3Sine}}
\end{figure}

As discussed in Ref.~\cite{Orford20}, the most evident systematic effect is a clear $t_{\text{acc}}$-dependence to the measured $\nu_c$ that is described as a sinusoidal oscillation at the magnetron frequency around the true $\nu_c$; because of this, measurements of the $\nu_c$ were taken across approximately one and a half magnetron periods and fit with the model described in Ref.~\cite{Orford20} to produce the true $\bar{\nu}_c$. For the case of low-yield $^{133m}\text{I}^+$ two measurements were taken at a half-magnetron-period of difference and a weighted average was calculated instead. 
A sample plot of the eleven final phase measurements of $^{133g}\text{Te}^+$  with $t_{\text{acc}}$ between 450.013 ms and 451.096 ms and the fit for the true cyclotron frequency $\bar{\nu}_c$ can be seen in Fig. \ref{fig:3Sine}. 
Additional possible systematic uncertainties associated with temporal instabilities in the magnetic field, the electric field in the Penning trap, ion-ion interactions, and the non-circular final projection of the phase have been studied~\cite{Ray24} and determined to have a cumulative effect smaller than 4 parts in $10^9$; this was added in quadrature. 

To determine the mass of isotope $X$, the magnetic field strength was then calibrated with the well-known mass of $^{133}\text{Cs}^+$~\cite{AME20}, and then the ratio of these two frequencies ${R=\bar{\nu}_c(^{133}\text{Cs}^+)/\bar{\nu}_c(^{133}\text{X}^+)}$ is used,
\begin{equation}
    M(^{133}\text{X})=[M(^{133}\text{Cs})-m_e]R+m_e,
\end{equation}
where $m_e$ is the electron mass. 
A summary of masses measured in this work and a comparison with previously measured values is presented in Table~\ref{tab:mass}.

\section{Results}
\subsection{$^{133}$Sb}
The result for $^{133}$Sb is in very good agreement with the Penning trap measured masses by Hakala {\it et al.}~\cite{Hakala12} and Van Schelt {\it et al.}~\cite{VanSchelt13}, as well as with the recommended AME2020 value, but it is better than a factor of two more precise. 
Due to its shorter half-life, the 16.54(19) $\mu$s isomeric state in $^{133}$Sb~\cite{NUBASE20} could not be seen.

\subsection{$^{133g,m}$Te}
The mass excess of $^{133g}$Te has a $\sim2\sigma$ tension with the value reported by Hakala {\it et al.}~\cite{Hakala12} and it is in a severe disagreement with the previous CPT result by Van Schelt {\it et al.}~\cite{VanSchelt13}. 
The present value for the $^{133m}$Te isomer is in good agreement with the direct mass measurement of this state by Kankainen {\it et al.}~\cite{Kankainen13} and with the previous CPT result by Van Schelt {\it et al.}~\cite{VanSchelt12} but is a factor of three more precise. 
For production of $^{133}$Te from a spontaneously-fissioning $^{252}$Cf source, we observed that the higher-spin isomeric state was more abundant than the ground state, which could explain why previous results are in better agreement with each other and with our new result for the isomer than for the ground state. 
Overall, the present results for $^{133g,m}$Te are factor of 3 more precise than the values recommended in AME2020. 
From the presently measured $^{133g,m}$Te masses, the excitation energy of the $^{133m}$Te isomer is determined as $E_{m}$=336.2(10) keV, which is close to the more precise NUBASE2020 value of 334.26(4) keV.   

\subsection{$^{133g,m}$I}
The mass excess of $^{133g}$I is a factor of six more precise and shows a $\sim2\sigma$ tension with the previously reported CPT results by Van Schelt {\it et al.}~\cite{VanSchelt13}, as well as with the AME2020 recommended data. 
The value for $^{133m}$I has the same precision as that in Ref.~\cite{NUBASE20}, and the deduced excitation energy of $E_{m}$=1636.0(58) keV is consistent with the recommended high-precision excitation energy of 1634.148(10) keV~\cite{NUBASE20}. 

\section{Discussion}
The masses of $^{133g}$Te and $^{133g}I$ measured in this present work can be used to determine new $Q_{\beta^-}(^{133}\text{Te})=2934.8(11)$ keV, which is about 16$\sigma$ higher and a factor of 6 more precise compared to the AME2020 value.
These new data resolve the $Q_{\beta^{-}}^{i}$ anomaly at the $E_{i}$=2935.83(15) keV level in $^{133}$I, resulting in the revised $Q{_{\beta^-}^i}=-1.0(12)$ keV. 

This offers the possibility that this state might be an ultralow Q-value decay. Such decays, which occur to excited states where $Q_\beta^i<1$~keV, are possible, and are of interest to neutrino physics research and to the study of atomic interference effects in the nuclear decay processes \cite{Keblbeck23,Mustonen10}. Promising candidates, such as $^{115}$In \cite{CATTADORI05,Mount09,Wieslander09}, $^{159}$Dy \cite{Ze21}, and $^{131}$I \cite{Eronen22} have been identified. It should be noted that due to its short half-life, a neutrino mass measurement using $^{133}$Te would not be able to rely on conventional production and supply methods (including isotope harvesting) \cite{Keblbeck23}. However, such a measurement could be done at a radioactive beam facility that providing parasitic beams directly to the experimental setup for sufficiently long periods of time to perform the measurement. Such a measurement presents several potential advantages. Unlike many ultralow Q-value decays, $^{133}$Te benefits from a short half-life and reasonably large branching ratio, resulting in an enhanced number of events near the decay end-point energy. Nevertheless, prior to considering such an effort, a mass measurement of $^{133}$Te and $^{133}$I with a precision better than 100 eV would be needed to pinpoint the Q-value.

It should also be noted that using $Q_{\beta^-}(^{133}\text{Te})<2935.9$ keV, determined from the present data, and the $\beta^{-}$-decay branching intensity reported in Ref. \cite{Khazov11} of 0.129(33)\% to the 2935.83(15) keV level, an anomalously-small $ft<3\times10^{-6}$ value to this level can be obtained using the LOGFT code~\cite{Gove71}. 
Similarly, unusually small $ft$ values can also be determined for several other highly-excited levels in $^{133}$I using the branching intensities of Ref.~\cite{Khazov11}.
These seemingly unphysical $\beta$-decay transition strengths indicate flaws in our current understanding of the structure of this nucleus.
Thus, future $\beta^{-}$-decay spectroscopy studies of $^{133g,m}$Te are necessary in order to confirm their decay schemes and branching intensities. 

Nuclear isomers in this area are also of interest to the study of the astrophysical r-process, particularly those that are thermally populated during the r-process and have a significantly different half-life \cite{Misch21}. $^{133m}$Te has been identified as one such ``astromer'', delaying $\beta$ decay \cite{Misch21} versus the ground state in the first hour of the r-process and impacting the formation of the $N=82$ abundance peak \cite{Misch21b}. However, the impact and population of an astromer depends on transitions through the intermediate nuclear states \cite{Misch21b}. Given the anomalies identified above, future studies of the structure of $^{133}$Te might also impact its role as an astromer in the $r$-process.

\section*{Conclusion}
In summary, high-precision mass measurements of $^{133}$Sb, $^{133g,m}$Te, and $^{133g,m}$I are reported, allowing for a calculation of $Q_{\beta^-}(^{133}\text{Te})$ to a precision of 1.1 keV, a sixfold improvement over the previous value, and which resolves an anomalous decay to the $E_i=2935.83(15)$ keV level of $^{133}$I as $Q^i_\beta=-1.0(1.2)$ keV. Further problems with the nuclear structure are indicated, as anomalously small $ft$ values for this and several other transitions can be calculated, so $\beta^-$-decay spectroscopy studies are needed. Such studies will further elucidate the potential of the decay to this excited state to be an ultra-low Q-value decay, and the role of $^{133m}$Te as an astromer.

\section*{Acknowledgments} 
This work was performed with the support of US Department of Energy, Office of Nuclear Physics under Contract No. DE-AC02-06CH11357 (ANL), the Natural Sciences and Engineering Research Council of Canada under Grant No. SAPPJ-2018-00028, and the US National Science Foundation under Grant No. PHY-2310059. This research used resources of ANL’s ATLAS facility, which is a DOE Office of Science User Facility. 

\bibliographystyle{elsarticle-num} 
%\bibliography{A133} 

\end{document}